# Correlation of high-field and zero- to ultralow-field NMR properties using 2D spectroscopy

Ivan V. Zhukov,[a,b] Alexey S. Kiryutin,[a,b] Alexandra V. Yurkovskaya,[a,b]

John W. Blanchard,[c,d,e] Dmitry Budker,[d,e,f] Konstantin L. Ivanov[a,b,*]

[a] *International Tomography Center, Novosibirsk, 630090, Russia*

[b] *Novosibirsk State University, Novosibirsk, 630090, Russia*

[c] *NVision-Imaging Technologies, Ulm, 89081, Germany*

[d] *Johannes Gutenberg-Universität Mainz, 55128 Mainz, Germany*

[e] *Helmholtz-Institut, GSI Helmholtzzentrum für Schwerionenforschung, 55128 Mainz, Germany*

[f] *Department of Physics, University of California, Berkeley, California 94720, USA*

\* Corresponding author, email: ivanov@tomo.nsc.ru

## Abstract

The field of zero- to ultralow-field (ZULF) nuclear magnetic resonance (NMR) is currently experiencing a rapid growth, owing to the progress in optical magnetometry, and also attractive features of ZULF NMR, such as low hardware cost and excellent spectral resolution achieved under ZULF conditions. In this work, an approach is proposed and demonstrated for simultaneous acquisition of ZULF-NMR spectra of individual $^{13}$C-containing isotopomers of chemical compounds in a complex mixture. The method makes use of fast field cycling, so that the spin evolution takes place at ZULF conditions, whereas signal detection is performed in a high-field NMR spectrometer. This method has excellent sensitivity, also allowing easy assignment of ZULF-NMR spectra to specific analytes in the mixture. We demonstrate that the spectral information is the same as that given by ZULF-NMR, which makes the method suitable for creating a library of ZULF-NMR spectra of various compounds and their isotopomers. The results of the field-cycling experiments can be presented in a convenient way as 2D-NMR spectra with the direct detection giving the high-field $^{13}$C-NMR spectrum (carrying the chemical-shift information) and the indirect dimension giving the ZULF-NMR spectrum (containing information about proton-carbon J-couplings). Hence, the method can be seen as a variant of heteronuclear J-resolved spectroscopy, one of the first 2D-NMR techniques.

## I. Introduction

Zero- to ultralow-field (ZULF) variant[1-5] of nuclear magnetic resonance (NMR) spectroscopy is drawing increasing attention in the last years. This approach is different from conventional NMR spectroscopy, in which it is desirable to work at highest available field in order to increase spectral resolution and sensitivity. In conventional NMR, to measure the signal, spins are excited with a radiofrequency pulse and their response is obtained by using inductive detection. By contrast, in ZULF NMR the external field is strongly reduced (in most cases, reduced as much as possible) and the response of the spin system is detected using sensitive superconducting interferometers or atomic magnetometers[6], since inductive measurements under such conditions are extremely inefficient. ZULF NMR can be a useful alternative to traditional NMR methods because of its lower cost and smaller size of the spectrometer. Furthermore, ZULF NMR is not sensitive to variation of magnetic susceptibility across the sample, also being applicable to samples enclosed in metal containers[7, 8]. Of course, ZULF-NMR spectra do not carry any information about chemical shifts of the nuclear resonance, but they are usually characterized with excellent resolution and allow one to determine spin-spin interactions, i.e., scalar J-couplings[1-3, 9]. Such ZULF J-spectra can be obtained for heteronuclear spin systems, e.g., those containing protons and $^{13}$C nuclei.

Despite these remarkable features of ZULF NMR, analysis of J-spectra meets difficulties, requiring simulations of the spin dynamics of a strongly coupled spin system. When the sample contains several analytes, one more problem is assignment of spectral peaks to specific chemical species in the sample. Solution of the assignment problem would then require spin-dynamics simulations of the molecules of interest or creating a library of their ZULF-NMR spectra recorded experimentally. Creating such a library can be time consuming, because of the need for measuring ZULF-NMR spectra for multiple samples containing a single analyte and also because the sensitivity of ZULF-NMR is still much lower than that of the conventional high-field NMR methods. The problem is further aggravated for samples with natural isotopic abundance: in this case several isotopomers of each analyte are present at a lower concentration, each having a different J-spectrum. For instance, at natural isotope abundance for ethanol there are two different isotopomers containing a single $^{13}$C atom with the concentration roughly 100 times lower than that of the unlabeled compound (the concentration of the isotopomer with two $^{13}$C nuclei is tiny and can usually be neglected for unlabelled samples).

In this work, we describe an approach, which allows one to perform fast acquisition of multiple ZULF-NMR spectra for all individual chemical species (and even their isotopomers) in the sample. The approach is using a combination of high-field (HF) NMR and ZULF NMR and allows one to correlate HF and ZULF-NMR spectra of the studied compounds. For convenience, we present the results in the form of a 2D-NMR spectrum, in which the direct dimension gives the HF spectrum and the indirect dimension gives the ZULF spectrum. The experiments have been carried out for a sample containing several small molecules with natural isotopic abundance. We have recorded HF spectra for all $^{13}$C nuclei (1.1% natural abundance), whereas ZULF-NMR spectra for each isotopomer of each compound are given by the J-couplings in the corresponding molecule. To facilitate peak assignment, we propose here to record $^{13}$C HF-NMR spectra with proton decoupling. In this situation, each $^{13}$C peak in the direct spectral dimension corresponds to a specific molecule (the presence of two or more $^{13}$C nuclei in one molecule is highly improbable), whereas the indirect dimension gives the sought after ZULF-NMR spectrum of this particular analyte. Such an approach, combing HF and ZULF properties, thus differs from 2D-ZULF-NMR methods introduced earlier.[10, 11] This kind of 2D-NMR spectroscopy can also be considered as a variant of J-resolved spectroscopy,[12-14] in which heteronuclear J-couplings are mapped out.

## II. Methods

### A. Materials

The experiments were carried out with a sample made by mixing equal volumes of methanol ($CH_3OH$, HPLC-grade, Chimmed), 95% ethanol ($C_2H_5OH$, OAO Kemerovskaya farmacevticheskaya fabrika), acetonitrile ($CH_3CN$, HPLC-grade, Chimmed), glacial acetic acid ($CH_3COOH$, Chimmed), and deuterated dimethyl sulfoxide (DMSO-$d_6$, Deutero GmbH), the latter added for long-term stabilization of the spectrometer magnetic field using the deuterium lock signal. The final concentrations of the substances in the sample are: methanol – 4.94 M, ethanol – 3.26 M, acetonitrile – 3.85 M, acetic acid – 3.5 M. Except for DMSO-$d_6$, substances with natural he abundance were used, giving rise to several mM concentrations of $^{13}$C-containing isotopomers. Structures of five $^{13}$C-isotopomers (with proton-carbon J-couplings specified), along with the $^1$H and $^{13}$C NMR spectra of the mixture, are shown in **Figure 1**.

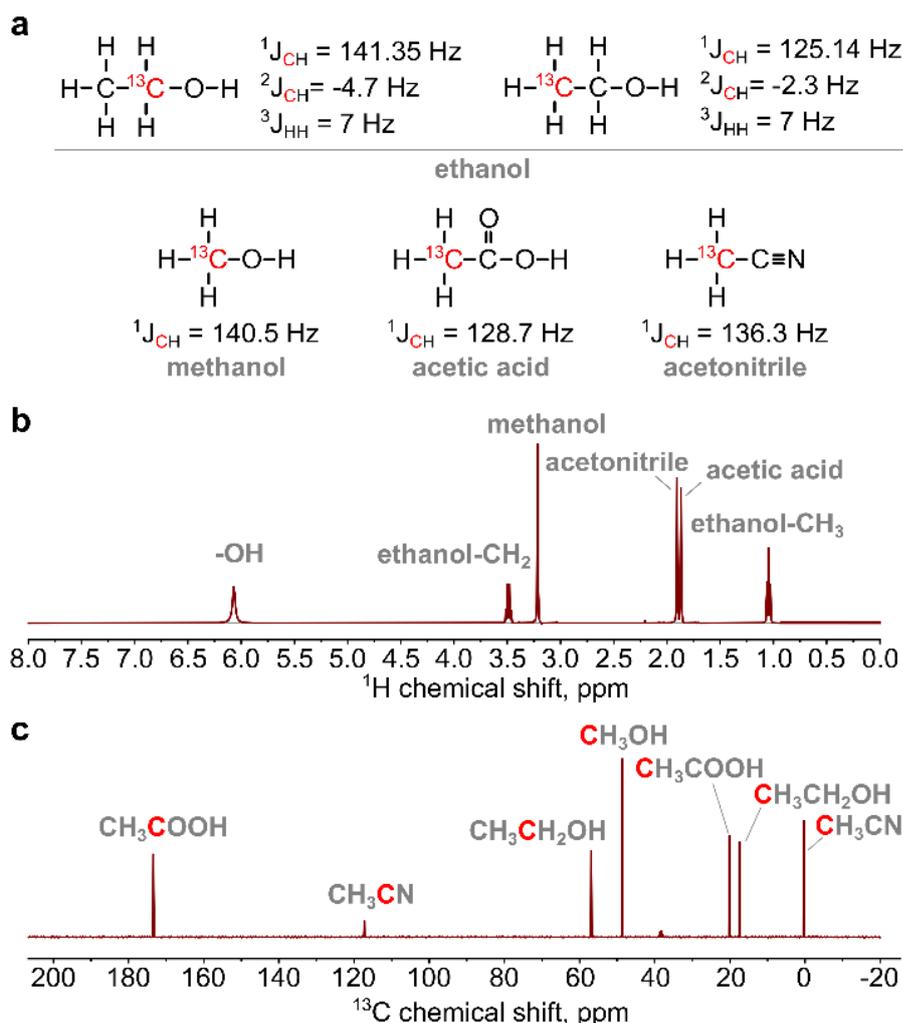

**Figure 1.** (a) Structures of the five $^{13}$C-isotopomers studied in the work; proton-carbon J-couplings are also indicated. Substances with natural isotopic distribution were used to prepare this mixture. (b) 400.13 MHz $^1$H NMR spectrum of the mixture. The signals in the spectrum predominantly come from $^{12}$C-isotopomers. (c) 100.62 MHz $^{13}$C NMR spectrum of the mixture ($^1$H-decoupled). Two signals at the low-field part of the spectrum correspond to the $^{13}$C-carbonyl acetic acid (173.4 ppm) and $^{13}$C-cyano acetonitrile (117.3 ppm) isotopomers, having a much weaker coupling to protons than the directly bound carbons. The $^{13}$C NMR signal of DMSO-d$_6$ at 38 ppm is broader than other signals due to the multiplet structure caused by scalar coupling to the six deuterium nuclei.

## B. Field-cycling NMR

To run field-cycling NMR experiments we used a setup[15, 16] for mechanical positioning of the NMR sample in an inhomogeneous magnetic field. The sample is shuttled along the NMR bore axis in the stray field of the NMR magnet. To extend the accessible field range to the μT and sub-μT range, a magnetic shield was mounted on top of the NMR magnet. Inside the shield, a magnetic coil system was introduced, allowing one to vary the field from the lowest field of about 5 nT to the field of the order of 1 mT. To shuttle the sample, a computer-controlled step motor was used, which provides fast and reproducible sample transport. The time of transporting the sample from the NMR spectrometer to the magnetic shield was 0.4 s. Further information about the setup can be found in previous publications[15, 16].

The idea of ZULF-NMR experiments is illustrated in **Figure 2**, with the experimental protocol depicted in subplot (a). The protocol comprises five consecutive steps. In step (i) spins acquire their equilibrium polarization at the $B_0$ field of an NMR spectrometer. Since equilibrium polarization is proportional to the Zeeman interaction of the corresponding spin (which is, in turn, proportional to its gyromagnetic ratio), heteronuclei acquire different equilibrium magnetization. For instance, the density matrix of a heteronuclear *IS* spin system, here an ¹H-¹³C spin system, is given by [17]:

$$\rho \propto M_I \hat{I}_z + M_S \hat{S}_z, \qquad (1)$$

where $M_I = \frac{\hbar \gamma_I B_0}{2k_B T}$ and $M_S = \frac{\hbar \gamma_S B_0}{2k_B T}$ stand for the equilibrium polarizations at the $B_0$ field ($\gamma_{I,S}$ are the nuclear gyromagnetic ratios, $k_B$ is the Boltzmann constant and $T$ is the absolute temperature). In this expression we omit the unity operator, as it is usually done in NMR. Since $\gamma_I \neq \gamma_S$ we immediately obtain $M_I \neq M_S$. The eigenstates of the spin system are the Zeeman states $|\alpha\alpha\rangle$, $|\alpha\beta\rangle$, $|\beta\alpha\rangle$ and $|\beta\beta\rangle$, with $|\alpha\rangle$ and $|\beta\rangle$ denoting the "spin-up" and "spin-down" states, respectively. When $M_I \neq M_S$, the states $|\alpha\beta\rangle$ and $|\beta\alpha\rangle$ acquire different populations and the population difference $\delta P$ of these states is directly proportional to $(M_I - M_S)$.

After preparation, a magnetic field jump is performed and the system is brought to an ultralow field $B_{UL}$ in a non-adiabatic fashion, meaning that the spin state does not change during the field jump, see step (ii). To implement such a field jump experimentally[15, 16, 18] we proceed as follows. First, the sample is brought to a small residual field $B_L$ of 26 µT, in which the spins are still coupled weakly so that the eigenstates do not alter. After that the field is rapidly reduced to $B_{UL}$ of about 100 nT by switching the current in a magnetic coil. At the $B_{UL}$ field the spins are coupled strongly and the eigenstates of the spin system are the singlet-triplet states, $|T_M\rangle$ and $|S\rangle$:

$$|T_{+1}\rangle = |\alpha\alpha\rangle, \quad |T_0\rangle = \frac{|\alpha\beta\rangle + |\beta\alpha\rangle}{\sqrt{2}}, \quad |T_{-1}\rangle = |\beta\beta\rangle; \quad |S\rangle = \frac{|\alpha\beta\rangle - |\beta\alpha\rangle}{\sqrt{2}}. \qquad (2)$$

If the density matrix of the spin system does not change during the field jump, the population difference $\delta P$ is converted into zero-quantum coherence (ZQC) $\rho_{ST_0}$ between the $|S\rangle$ and $|T_0\rangle$ states[16, 19, 20]. Hence, during the free evolution period of a duration $\tau_{ev}$, which constitutes step (iii), the time dependence of the density matrix is due to the evolution of $\rho_{ST_0}$, which is coherent oscillations at the frequency $J_{IS}$.[19, 20] For detecting spin magnetization, two more steps are required. Step (iv) is a two-stage non-adiabatic field jump $B_{UL} \to B_L \to B_0$, which converts the $\rho_{ST_0}$ ZQC back into the population difference $\delta P$ between the states $|\alpha\beta\rangle$ and $|\beta\alpha\rangle$, corresponding to the difference $(M_I - M_S)$ in the longitudinal spin polarizations. Finally, a hard 90° radiofrequency (RF) pulse is applied and in step (v) the NMR spectrum of either spin *I* or spin *S* is taken.

As was shown previously,[19, 20] in this experiment the $M_{I,S}$ polarizations oscillate as a function of the evolution time and so does the $\delta P$ value:

$$(M_I - M_S) \propto \delta P \propto \cos[2\pi J_{IS} \tau_{ev}]. \qquad (3)$$

The $(M_I - M_S)$ behavior can be monitored directly at the ultralow field[6] by using an atomic magnetometer or at high field[16, 19, 20] by repeating the field-cycling experiment multiple times with incremented $\tau_{ev}$. In the latter case, one can monitor separately either $M_I(\tau_{ev})$ or $M_S(\tau_{ev})$ by using NMR detection at the corresponding channel. Both experiments, detection at the ultralow or high field, are expected to give the same time traces, as demonstrated in this work. Here we mostly use the latter strategy and analyze $M_I$ or $M_S$ as a function of $\tau_{ev}$. In most cases, it is preferable to measure such curves for ¹³C spins, because for protons the NMR lines of the relevant isotopomers are satellites of much more intense lines of the molecules containing no ¹³C nuclei.

A typical dependence of the ¹³C signal on $\tau_{ev}$ is depicted in **Figure 2b**. In this example, we show the results for methanol containing a ¹³C label in the methyl group. One can clearly see fast oscillations predominantly given by the proton-carbon $J_{IS} = J_{CH}$ coupling in the methyl group. Additionally, relaxation effects come into play (which were not discussed above) resulting in decay of the signal and damping of the oscillations. The decaying signal component (averaged over oscillations) was subtracted from these curves prior to applying Fourier transformation, which gives J-resolved spectrum of the system at the $B_{UL}$ field, i.e., the ZULF-NMR spectrum. As we show below, such a spectrum contains the same information as the one taken directly at the $B_{UL}$ field. In more complex cases, where the spin system contains more than two nuclei, one can detect ZULF-NMR spectra, which contain peaks at characteristic frequencies, reflecting the J-coupling topology. Generally speaking, in multiple-spin systems one should run numerical simulations of the spin dynamics in order to analyze such J-spectra.

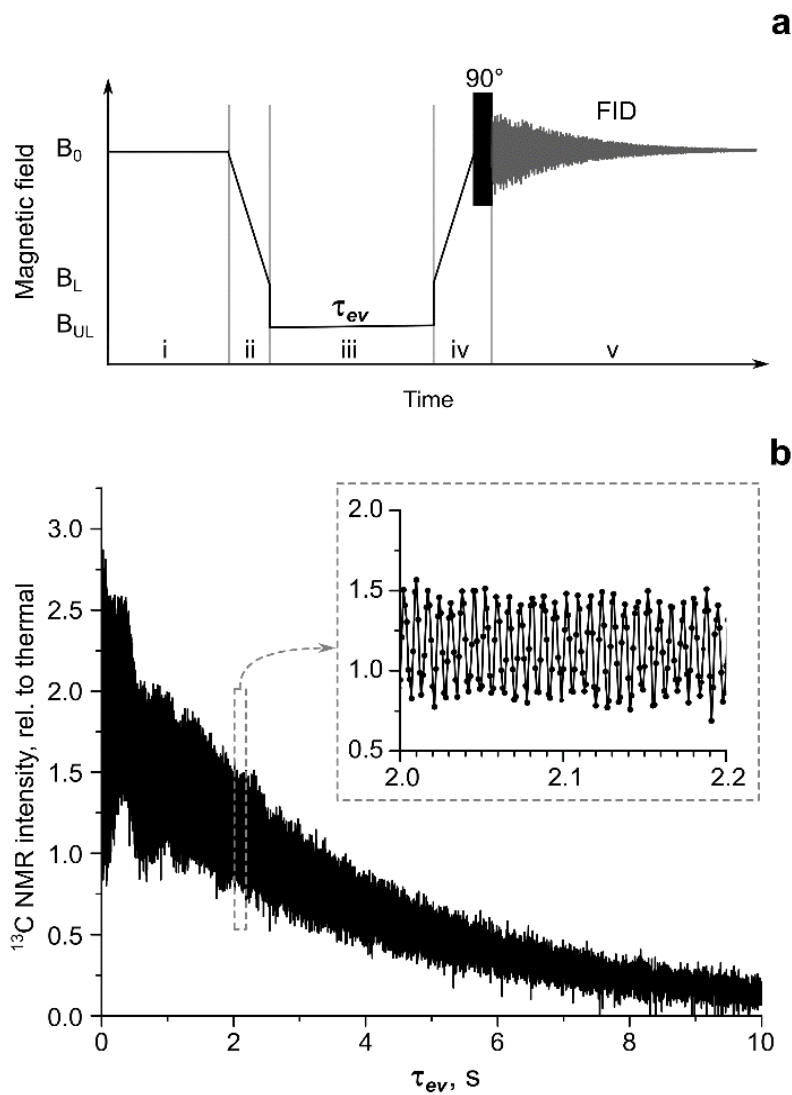

**Figure 2**. (a) Protocol of a field-cycling experiment, here used to measure the coherent spin evolution at ZULF conditions. The protocol comprises the following steps: (i) relaxation of the spin system to thermal equilibrium at the high magnetic field $B_0$; (ii) non-adiabatic field switching: mechanical sample transfer into the magnetic shield with a pre-set low magnetic field $B_L$ inside, which is non-adiabatically switched to an ultralow field $B_{UL}$ immediately after sample arrival. The field jump $B_L \rightarrow B_{UL}$ creates heteronuclear ZQC, which evolves in step (iii) during a free evolution period $\tau_{ev}$ of variable length; (iv) non-adiabatic field switching $B_{UL} \rightarrow B_L \rightarrow B_0$, followed by application of an RF pulse at the resonance frequency of the nuclei to be observed; (v) free induction decay acquisition. The NMR spectrum is given by the Fourier transform of the FID. (b) Dependence of methanol $^{13}$C NMR signal intensity on $\tau_{ev}$. Insert shows fast oscillations of the signal. Parameters of the experiment are: $B_L = 26\,\mu T$, $B_{UL} = 100\,nT$, 1 scan per point, relaxation delay 12.4 s, $\tau_{ev}$ incremented from 10 ms to 10.249 s with 1 ms step, total experiment duration is ca. 60 hours.

Finally, we would like to emphasize that the protocol shown in **Figure 2a** can be considered as a protocol of a 2D-NMR experiment. Indeed, steps (i) and (ii) can be seen as preparation of the spin order, step (iii) is free evolution of the coherence of interest (in this case, of the heteronuclear ZQC), whereas the field jump $B_{UL} \rightarrow B_0$ and the subsequent RF pulse perform mixing of the spin order, which is then detected by NMR. Hence, the building blocks of a 2D-NMR experiment, preparation-evolution-mixing-detection, are clearly present. Moreover, the time-domain FID signal, $S(t)$, has a parametric dependence on the $\tau_{ev}$ time. Hence, the signal depends on two times $\mathcal{S}(\tau_{ev}, t)$ and its Fourier transform yields a 2D-NMR spectrum $\mathcal{S}(f_1, f_2)$, where $t$ encodes the direct domain and $\tau_{ev}$ encodes the indirect domain. The spectrum in the direct domain is a HF-NMR spectrum with line positions corresponding to $^{13}$C chemical shifts, whereas in the indirect domain we obtain the ZULF-NMR spectrum, which contains multiple peaks due to J-couplings. As we show below, presentation of the results in the form of a 2D spectrum obviates assignment of ZULF-NMR peaks to specific isotopomers in the sample.

### C. ZULF NMR

In this work, we also compare ZULF-NMR spectra obtained using field cycling with those taken directly at an ultralow field. ZULF-NMR spectra have been obtained for isotopically-enriched neat liquids. Samples were pre-polarized at 1.8 tesla; ZULF-NMR signals were detected using sensitive optical magnetometers. Acquisition parameters and details of sample preparation are described in Refs. 21 and 22 for methanol and ethanol spectra, respectively.

## III. Results and Discussion

A 2D-NMR spectrum of the studied sample is shown in **Figure 3**. As expected, the spectrum in the direct domain is the $^{13}$C-NMR spectrum of the mixture of the compounds under investigation (when proton decoupling is used, this spectrum contains only $^{13}$C chemical shift information). The ZULF-NMR spectrum detected in the indirect domain is more complex, containing peaks at various frequencies, ranging from a few Hz to several 100 Hz. When $^{13}$C-signals are unresolved, assignment of these peaks is problematic: such a spectrum without $^{13}$C chemical shift resolution is shown in **Figure 4**.

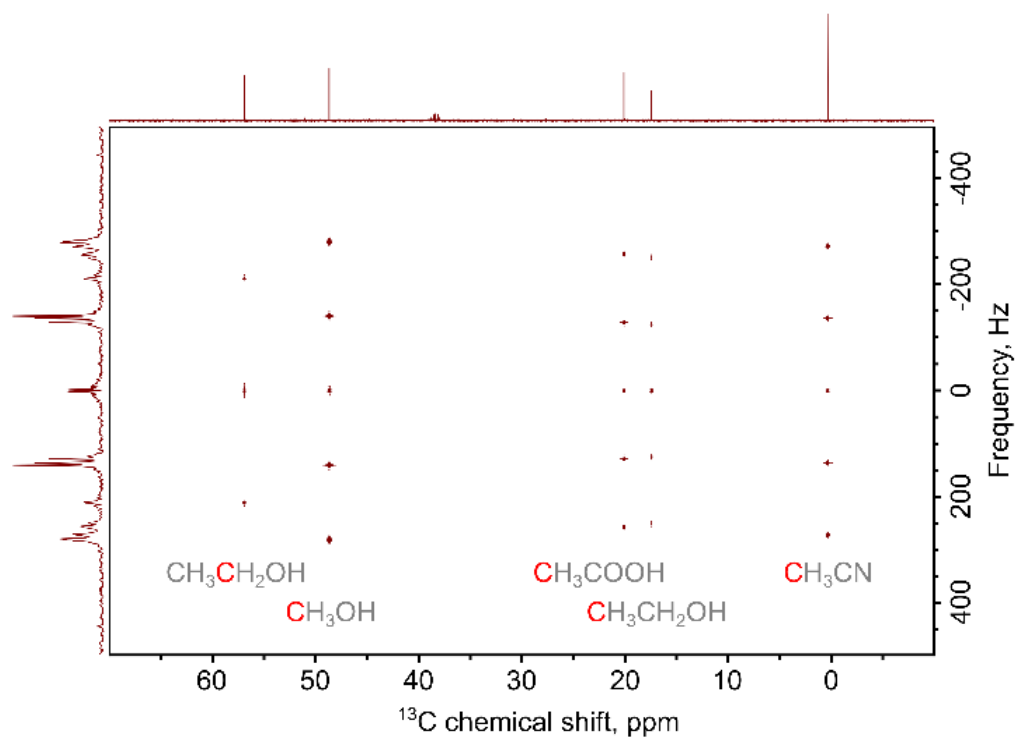

**Figure 3**. 2D HF-ZULF-spectrum of the mixture. In this spectrum, the horizontal axis represents chemical shift values of $^{13}$C NMR signals, while the vertical axis shows frequencies derived from the coherent spin evolution under ZULF conditions. The spectrum presents a magnitude-mode Fourier transform of the $\mathcal{S}(t, \tau_{ev})$ signal, shown by 1500×4096 points. Apodization of the acquired data with decaying exponential functions (10 Hz along the horizontal axis and 1 Hz along the vertical axis) was applied prior to the 2D Fourier transform. One-dimensional spectrum on the left shows the projection of the 2D spectrum on the indirect domain, i.e., the ZULF-NMR spectrum. One-dimensional spectrum on top shows the standard $^{13}$C NMR spectrum acquired with $^{1}$H decoupling.

The assignment problem can be solved by using 2D-NMR spectra. Indeed, chemical shift resolution provided by HF-NMR immediately provides correlation between the HF and ZULF properties. Hence, for each chemical shift we obtain a resolved J-coupling spectrum. For this reason, one can interpret this method as a variant of J-resolved spectroscopy[12-14].

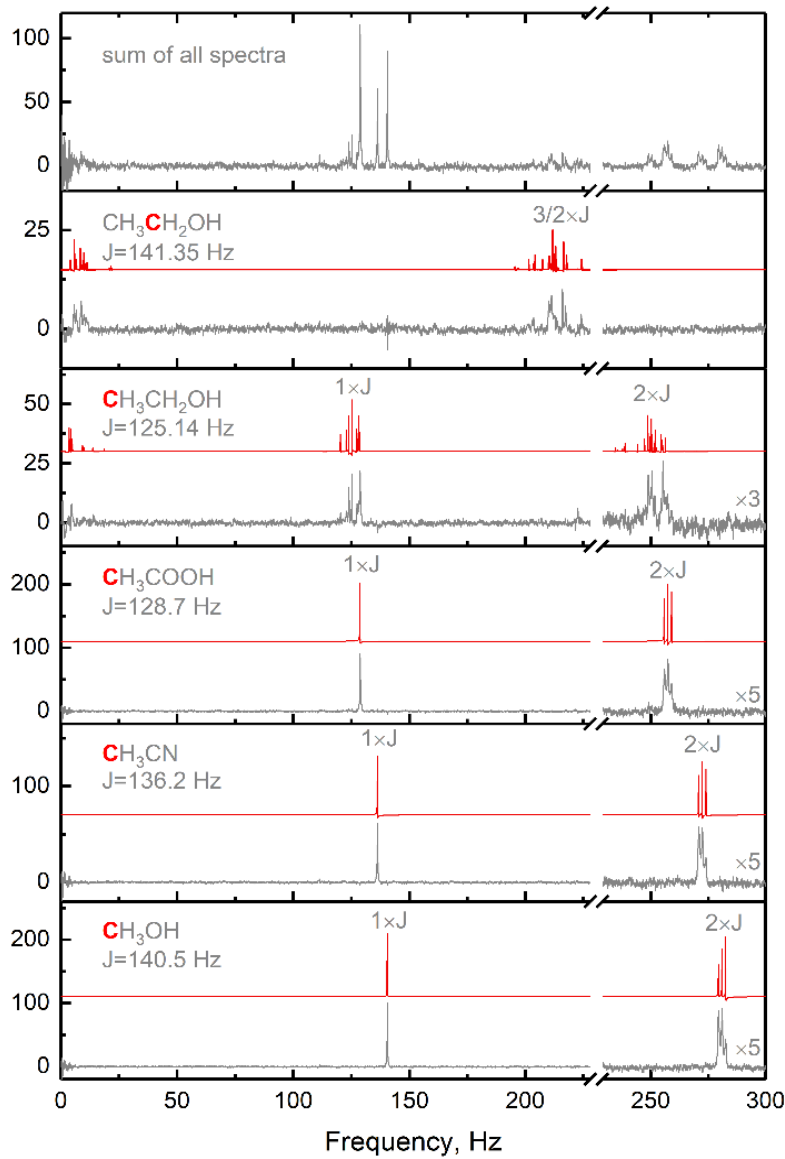

**Figure 4**. Measured (grey lines) and calculated (red lines) ULF-spectra of the individual compounds in the mixture. The spectrum at the top was obtained as a sum of the ZULF-spectra for all components.

In this context, it is also useful to present J-spectra of individual compounds and to compare them with simulation results. Such a comparison is given in **Figure 4**; the simulation was performed using software available online[23]. One can see that the simulation matches the experimental J-spectra well. For each compound, one can see pronounced peaks at frequencies in the range 120-300 Hz, coming from the interaction with the nearby protons. In the case of a single neighboring proton there is only one such peak at the frequency of $J_{CH}$. For a CH$_2$ group also a single peak at $\frac{3}{2}J_{CH}$ is expected, whereas for a CH$_3$-group there should be two peaks at $J_{CH}$ and $2J_{CH}$, in accordance with ZULF-NMR theory.[1] One should also note that peaks can be additionally split due to smaller J-couplings with other protons in the molecule and also due to the Zeeman interactions with residual magnetic field, which is $B_{UL} = 100$ nT.[1] Additional structure is indeed seen for the compounds and it is reproduced by the simulation.

We point out that spectra of the isotopomers containing a $^{13}$C-labeled methyl group are very similar, due to the presence of intense peaks at $J_{CH}$ and $2J_{CH}$ and also due to the similar values of the $J_{CH}$ coupling in the compounds under study. For these reasons, analysis of ZULF-NMR spectra and assignment of the spectral peaks might be complicated, whereas HF detection can easily resolve this problem, making assignment straightforward.

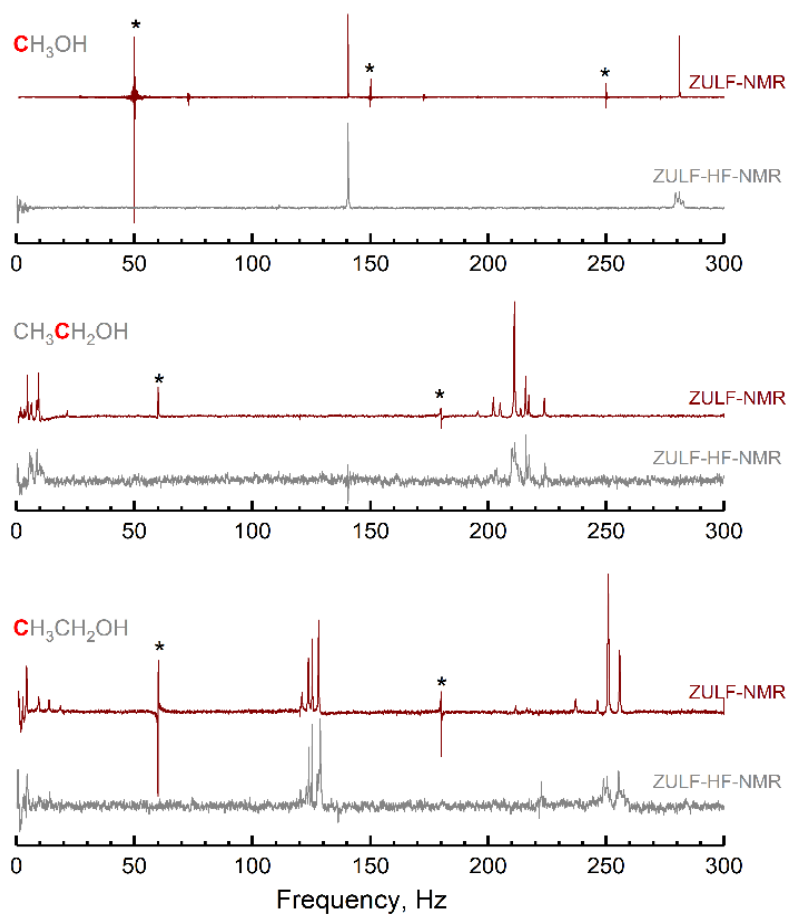

**Figure 5**. Comparison of J-spectra measured by optical magnetometry at zero field and using field-cycling and high-field NMR detection. Spectra shown in grey (ZULF-NMR spectra) are the spectra of $^{13}$C-enriched samples detected using a ZULF-NMR spectrometer. Spectra shown in brown (ZULF-HF-NMR spectra) have been obtained in this work for the mixture of substances with natural isotopic abundance. To obtain ZULF-HF-NMR spectra field cycling was used, the $B_{UL}$ field was about 100 nT. Asterisks in ZULF-NMR spectra indicate artefacts coming from AC line.

Finally, it is useful to compare ZULF-NMR spectra obtained with field cycling with those obtained at ZULF conditions, using an optical magnetometer. Such a comparison is given in **Figure 5**, presenting the spectra of methanol and two isotopomers of ethanol. The spectra obtained using field cycling and high-field detection are denoted as "ZULF-HF" spectra, whereas the ZULF-NMR spectra obtained at an ultralow field are named "ZULF" spectra. One can see that such spectra match each other. Subtle differences come from the fact that in field-cycling experiments the $B_{UL}$ field was nonzero, which gave rise to additional splitting of the NMR lines. For instance, the line at $2J_{CH}$ in methanol is affected by the presence of the $B_{UL} \neq 0$ field. In the ZULF spectra, one can see additional artefacts, coming from AC line (ZULF-HF spectra are free from these artefacts). The ethanol spectra have been taken in Mainz (the AC line is at 50 Hz), whereas the methanol spectrum has been acquired in Berkeley (the AC line is at 60 Hz). Apart from this, good agreement of the ZULF-NMR and ZULF-HF-NMR spectra demonstrates that one can indeed use the method introduced here for obtaining ZULF-NMR spectra of various compounds.

Hence, the two methods for obtaining ZULF-NMR spectra match each other, each having its advantages and disadvantages. The approach using field-cycling also has a disadvantage that the evolution under ZULF conditions is measured point-by-point for each $\tau_{ev}$ value. At first glance, this experiment is excessively time consuming, since the result for every $\tau_{ev}$ value comes from a separate experiment cycle. Nonetheless, this disadvantages are minimized due to higher sensitivity and ease of signal assignment, making it possible to analyze simultaneously multiple compounds in the sample. Potentially, using $^{13}$C-NMR detection, one should be able to analyze dozens of analytes in the sample. HF detection also minimizes the sensitivity issue, allowing one to work with samples with natural isotopic abundance.

## IV. Conclusions

In this work, a method is proposed and demonstrated to obtain ZULF-NMR spectra of complex mixtures in a manner allowing one decomposition of the ZULF-NMR spectrum into spectra of individual compounds and, moreover, easy assignment of the individual spectra to specific isotopomers of each analyte. This is achieved by running a 2D-NMR experiment, in which the direct dimension is recorded at the high field of an NMR spectrometer and the indirect dimension is encoded by the spin evolution under ZULF conditions. Such a detection method provides the same ZULF-NMR spectra as those recorded directly under ZULF-NMR conditions.

The proposed method has its advantages and weaknesses. A disadvantage is that acquisition of ZULF-NMR spectra requires mapping the indirect dimension point by point, whereas the conventional scheme allow measuring ZULF-NMR spectra in a single experiment. This weakness is compensated by the following advantages: (i) high detection sensitivity, (ii) possibility of simultaneous acquisition of multiple compounds (and their isotopomers), (iii) decomposition of the ZULF-NMR spectrum into individual spectra, (iv) simple spectral assignment. For this reason, the introduced method can be used to create a library of ZULF-NMR spectra of compounds of interest.

Presentation of the results in the form of a 2D spectrum is convenient, enabling straightforward correlation of ZULF-NMR spectra with specific compounds. Such a presentation, in fact, allows one to obtain heteronuclear J-resolved spectra, which is useful for analyzing complex mixtures of multiple compounds and their isotopomers.

## Acknowledgements

Support from the Ministry of Science and Higher Education of the Russian Federation (Grant No. 075-15-2020-779) is gratefully acknowledged. This project has support from the European Union's Horizon 2020 research and innovation programme under the Marie Sklodowska-Curie Grant Agreement No. 766402.

## Data availability

The data that support the findings of this study are available from the corresponding author upon reasonable request.